\documentclass[letterpaper,twocolumn,showpacs,
prb
,superscriptaddress,email
,showkeys%
]{revtex4}

\usepackage{soul}
\usepackage[pdftex]{graphicx}
\usepackage{amssymb}
\usepackage{mathrsfs}
\usepackage{amsmath,amsfonts,latexsym}
\usepackage{array,tabularx,color}
\usepackage{dcolumn} 
\usepackage[normalem]{ulem}
\usepackage{comment}
\usepackage{bm}

\usepackage{doi}
\usepackage{hyperref}

\hypersetup{
    colorlinks=true,
    linkcolor=magenta,
    citecolor=magenta,
    filecolor=magenta,
    urlcolor=red,
}

\newcommand{\redc}[1]{{\color{black}#1}}

\begin{document}

\title{Optical control of spin textures in quasi-one-dimensional polariton condensates}

\author{C. Ant\'on}
\affiliation{Departamento de F\'isica de Materiales, Universidad Aut\'onoma de Madrid, Madrid 28049, Spain}
\affiliation{Instituto de Ciencia de Materiales ``Nicol\'as Cabrera'', Universidad Aut\'onoma de Madrid, Madrid 28049, Spain}

\author{S. Morina}
\affiliation{School of Physical and Mathematical Sciences, Nanyang Technological University, 637371, Singapore}

\author{T. Gao}
\affiliation{FORTH-IESL, P.O. Box 1385, 71110 Heraklion, Crete, Greece}
\affiliation{Department of Materials Science and Technology, Univ. of Crete, 71003 Heraklion, Crete, Greece}

\author{P. S. Eldridge}
\affiliation{FORTH-IESL, P.O. Box 1385, 71110 Heraklion, Crete, Greece}

\author{T. C. H. Liew}
\affiliation{School of Physical and Mathematical Sciences, Nanyang Technological University, 637371, Singapore}

\author{M. D. Mart\'in}
\affiliation{Departamento de F\'isica de Materiales, Universidad Aut\'onoma de Madrid, Madrid 28049, Spain}
\affiliation{Instituto de Ciencia de Materiales ``Nicol\'as Cabrera'', Universidad Aut\'onoma de Madrid, Madrid 28049, Spain}

\author{Z. Hatzopoulos}
\affiliation{FORTH-IESL, P.O. Box 1385, 71110 Heraklion, Crete, Greece}
\affiliation{Department of Physics, University of Crete, 71003 Heraklion, Crete, Greece}

\author{P. G. Savvidis}
\affiliation{FORTH-IESL, P.O. Box 1385, 71110 Heraklion, Crete, Greece}
\affiliation{Department of Materials Science and Technology, Univ. of Crete, 71003 Heraklion, Crete, Greece}

\author{I. A. Shelykh}
\affiliation{School of Physical and Mathematical Sciences, Nanyang Technological University, 637371, Singapore}

\author{L. Vi\~na}
\email{luis.vina@uam.es}

\affiliation{Departamento de F\'isica de Materiales, Universidad Aut\'onoma de Madrid, Madrid 28049, Spain}
\affiliation{Instituto de Ciencia de Materiales ``Nicol\'as Cabrera'', Universidad Aut\'onoma de Madrid, Madrid 28049, Spain}
\affiliation{Instituto de F\'isica de la Materia Condensada, Universidad Aut\'onoma de Madrid, Madrid 28049, Spain}

\date{\today}

\begin{abstract}
We investigate, through polarization-resolved spectroscopy, the spin transport by propagating polariton condensates in a quasi one-dimensional microcavity ridge along macroscopic distances. Under circularly polarized, continuous-wave, non-resonant excitation, a sinusoidal precession of the spin in real space is observed, whose phase depends on the emission energy. The experiments are compared with simulations of the spinor-polariton condensate dynamics based on a generalized Gross-Pitaevskii equation, modified to account for incoherent pumping, decay and energy relaxation within the condensate.
\end{abstract}



\maketitle

\section{Introduction}
\label{sec:intro}

Semiconductor microcavities (MCs) in the strong coupling regime are excellent candidates for designing novel ``spinoptronic" devices due to their strong optical nonlinearities,\cite{Tartakovskii:1999aa} polarization properties,\cite{Martin:2002aa,Kopotowski:2006aa,Kammann:2012aa} and fast spin dynamics\cite{Lagoudakis:2002aa}. The control of polariton condensates propagation and their polarization\cite{Kammann:2012aa} provide the necessary ingredients for future optical circuits. The first steps towards the fabrication of spin-based polariton condensate switches\cite{AmoA.:2010aa,Adrados:2011aa,PhysRevLett.112.046403,Grosso:2014aa}, gates\cite{PhysRevLett.99.196402} and memories\cite{Paraiso:2010aa,Cerna:2013aa} have been recently achieved. They fulfill the fundamental technological requirements for the operation with polarization-encoded signals: micrometric size, non-local action triggering and high-speeds (of the order of $\sim1$~$\mu$m/ps due to the ballistic polariton propagation). New schemes for the realization of spinoptronics devices\cite{PhysRevLett.100.116401,PhysRevB.81.125327} and ``polariton neurons" in circuits, the building blocks of all-optical integrated logic circuits,\cite{Liew:2008aa,PhysRevLett.102.046407,Espinosa-Ortega:2013aa} have been recently proposed. One-dimensional (1D) and quasi-1D patterned high-finesse MCs provide an ideal platform for all-optical manipulation,\cite{Wertz:2010ys} ballistic propagation and amplification\cite{Wertz:2012ee} and gating of polariton condensates.\cite{Gao:2012tg,anton:261116,Nguyen:2013aa,Anton:2013ab,Sturm:2014aa,Anton:2014aa} The waveguide nature of these structures induces the channeling of polariton propagation, while the discretization of energy levels results in a rich relaxation dynamics.\cite{Wouters:2012aa,Anton:2013aa}

\redc{In planar semiconductor MCs, the splitting of the transverse electric (TE) and magnetic (TM) modes of the cavity\cite{Panzarini:1999aa} induces an effective magnetic field, which on its own produces a precession of the polaritons spin, when they propagate over macroscopic distances. This effect is well-known as the optical spin Hall effect\cite{Leyder:2007aa,Kammann:2012aa} and it was first predicted by Kavokin and co-workers \cite{PhysRevLett.95.136601} as an analogue of the electronic spin Hall effect \cite{Dyakonov1971459,PhysRevLett.92.126603}. Initial experiments were conducted with resonant excitation~\cite{Leyder:2007aa} making use of Rayleigh scattering \cite{PhysRevLett.88.047401} or tightly focused laser spots \cite{PhysRevB.80.165325} to excite multiple states in reciprocal space. These experiments represented purely linear effects, not relying on the excitonic component of polaritons \cite{Maragkou:11}. The presence of quantum well excitons is required for non-resonant excitation, leading to the spontaneous formation of a propagating polariton condensate \cite{Kammann:2012aa}. The effective magnetic field representing the optical spin Hall effect can be utilized, for example, to generate polarization textures \cite{PhysRevB.75.075323,Kammann:2012aa}, where the polaritons propagate in rings spreading in real space, showing oscillations of the polarization degree in azimuthal angle and time; to convert the spin to orbital angular momentum \cite{PhysRevB.83.241307}; to create spin-polarized vortices \cite{PhysRevB.83.241307,Manni:2013aa,sala_arxiv14,PhysRevB.75.241301} and to form half dark solitons\cite{Hivet:2012aa,PhysRevB.83.193305} and very similar structures \cite{PhysRevLett.113.103901} in the wake of an obstacle. Recent theoretical work has also studied the role of the optical spin Hall effect in driving polarized bright solitons \cite{Egorov:14} and other spin patterns \cite{PhysRevB.90.165308,PhysRevB.89.235302}.}

In this work, we investigate optically the collective spin dynamics of polariton condensates moving along macroscopic distances in a quasi-1D MC ridge. \redc{The discretization in energy of the lower polariton branch (LPB) in our quasi-confined structure has notable consequences in the coherent transport of the spin vector. In the first place, the confinement renders a TE-TM mode splitting, which remains for zero in-plane wavevector, and acquires larger values than the TE-TM splitting in two-dimensional (2D) MCs. Furthermore, a spectral analysis of the spin transport reveals different polariton spin textures to those observed in 2D systems.\cite{Kammann:2012aa} The richness of these textures is related to the energy dependent speed of propagation of polaritons in our system with lowered dimensionality. The ballistic propagation of spin polarized polaritons along the ridge is observed over distances of $\sim100$ $\mu$m.}

To describe the polarization state of exciton-polaritons we adopt the pseudospin formalism.\cite{Kavokin:2004aa} Polaritons possess a spin with two possible projections on the structural growth axis of the MC. The polarization of the emitted light gives direct access to the pseudospin state, which is fully characterized by the four-component Stokes vector $\overrightarrow{s}=\left(s_0,s_x,s_y,s_z\right)$. Here, $s_0$ is the total photoluminescence (PL) intensity, and $s_{x,y,z}=\left(I_{H,D,\sigma^+}-I_{V,A,\sigma^-}\right)/\left(I_{H,D,\sigma^+}+I_{V,A,\sigma^-}\right)$. $I_{H,D,\sigma^+}$ and $I_{V,A,\sigma^-}$ are the measured intensities in the horizontal ($H$) and vertical ($V$), diagonal ($D$) and antidiagonal ($A$), and the two circular polarization components $\sigma^+$ and $\sigma^-$.

This paper is organized as follows. In Sec.~\ref{sec:sample_setup} we discuss the sample and the experimental setup. In Sec.~\ref{sec:exp} we present and discuss our results; we first show, under continuous wave (cw) excitation, the optical spin Hall effect \cite{PhysRevLett.95.136601,Kammann:2012aa} in a quasi-1D structure, discussing the $s_z$ oscillations in real space, for a $\sigma^+$-polarized pump. In Sec.~\ref{subsec:stokes} we systematically investigate the distribution of the Stokes components as function of the PL energy and position along the ridge. In Sec.~\ref{subsec:pump_pow} we demonstrate that the $s_z$ precession is lost under linear excitation and/or high power excitation conditions. In Sec.~\ref{sec:model} the experiments are compared with simulations of the spinor-polariton condensate dynamics based on a generalized Gross-Pitaevskii equation, modified to account for incoherent pumping, decay and energy relaxation within the condensate. Finally, in Sec.~\ref{sec:conclusions} we provide the conclusions of this work.
\begin{figure*}
\setlength{\abovecaptionskip}{-5pt}
\setlength{\belowcaptionskip}{-2pt}
\begin{center}
\includegraphics[width=0.95\linewidth,angle=0]{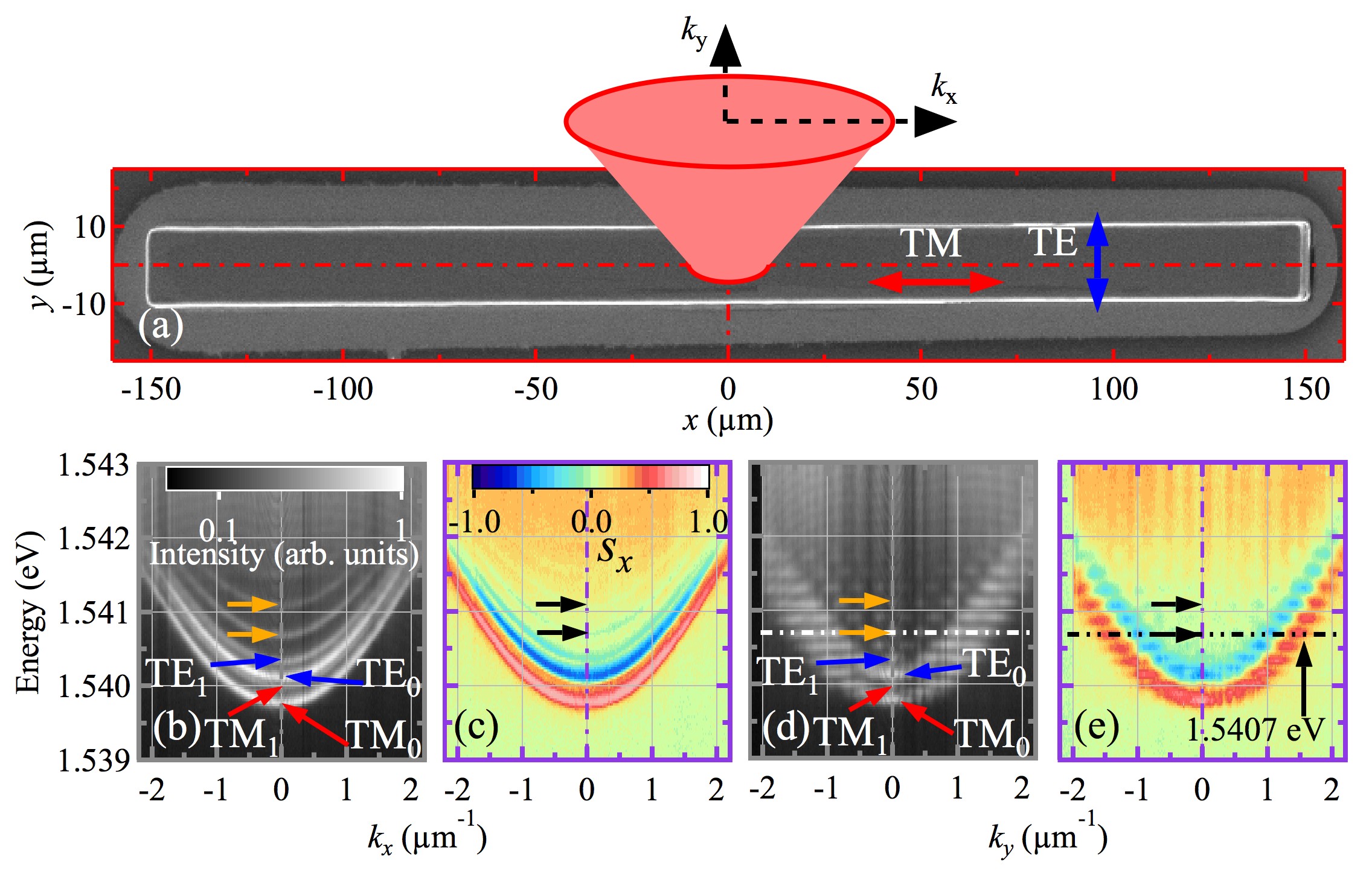}
\end{center}
\caption{(Color online) (a) Scanning electron microscopy image of a 20-$\mu$m wide ridge, including an angular scheme of the PL emitted from the center of the ridge (see cone of light as a guide to the eye), $z$ direction is perpendicular to the plane of the paper. The bottom panels display, under non-resonant (1.612 eV), weak, circularly polarized-light excitation: (b) and (d) energy dispersions of the PL along $k_x$ and $k_y$, respectively; (c) and (e) linear degree of polarization ($s_x$) versus energy and $k_x$ and $k_y$, respectively. Dot-dot-dashed white and black lines in panels (d,e) mark the energy value (1.5407 eV) used for Fig.~\ref{fig:dispersion2}(b). Red and blue arrows in panels (a,b,d) mark the TM and TE character of the even subbands in the dispersion relations, respectively. Orange and black arrows in panels (b-e) mark the energy positions of weakly-polarized, consecutive higher subbands. The PL and $s_x$ are coded in linear, normalized, false color scales.}
\label{fig:dispersion1}
\end{figure*}

\section{Sample and experimental setup}
\label{sec:sample_setup}

A high-quality $5\lambda/2$ AlGaAs-based MC with 12 embedded quantum wells is investigated, whose Rabi splitting, $\Omega_R$, amounts to $9$ meV. Ridges, with dimensions $20 \times 300~\mu$m$^2$, have been obtained by reactive ion etching (further information about this sample is given in Ref.~\onlinecite{Tsotsis:2012qy}). We study a ridge situated in a region of the sample corresponding to resonance (detuning between bare exciton and cavity modes is $\backsim$ 0). The sample is kept at 10 K in a cold-finger cryostat and it is excited with a cw laser, tuned to the first high-energy Bragg mode of the MC (1.612 eV). The cw laser is chopped at 300 Hz with an on/off ratio of 1:2 in order to prevent unwanted sample heating. We focus the laser beam on the sample through a microscope objective to form a 10~$\mu$m-$\varnothing$ spot. The same objective is used to collect (angular range $\pm18^\circ$) the PL, which is directed towards a 0.5 m imaging spectrometer. The power threshold for polariton condensation is $P_{th}=2$ mW.

In our experiments, polaritons propagate predominantly along the $x$ axis of the ridge [see Fig. \ref{fig:dispersion1}(a)]. Therefore in all the images presented in the manuscript, where the $y$ direction is not shown, the spectral PL distribution is analyzed along the $x$ axis from a $\Delta y=2$~$\mu$m-wide, central region of the ridge. However, for the sake of completeness, the full 2D polariton intensity and degree of circular polarization distributions are presented when appropriate.

We start by describing the dispersion relations of polaritons along two orthogonal directions in the ridge, $k_x$ at $k_y=0$ and $k_y$ at $k_x=0$. The confinement in the $y$ axis of the ridge results in the discretization of the $k_y$ in-plane momentum, splitting the LPB in many subbands, Fig.~\ref{fig:dispersion1}(b), whose antinodes along $k_y$ are visible in Fig.~\ref{fig:dispersion1}(d). It is important to emphasize that only even subbands are visible in Fig.~\ref{fig:dispersion1}(b), since along $k_x$ we spectrally resolve the PL at $k_y=0$. Odd modes (with a node at $\mathbf{k}=0$) are visible in Fig.~\ref{fig:dispersion1}(d), see for example the subband at 1.5405 eV. The scenery seen in these dispersion relations is very interesting because it reveals the possibility of parametric scattering processes among many different sub-branches. Recent works on 1D semiconductor MCs exploit these extra-confinement effects to study new parametric phenomena (see, for example, Refs. \onlinecite{PhysRevB.87.155302, PhysRevLett.108.166401} and references therein).

Confinement in a quasi 1D cavity enhances the splitting between the two light polarizations TM and TE, parallel and perpendicular to the $x$ axis of the ridge, respectively. We identify TM (TE) as the H (V) direction used to define the linear degree of polarization ($s_x$). Considering the $k_x$ direction, the lowest energy subband is TM$_0$ polarized, see label in Fig.~\ref{fig:dispersion1}(b), and its corresponding $s_x$ is shown in Fig.~\ref{fig:dispersion1}(c), where an intense, red subband appears, whose minimum is at $E=1.5397$~eV and $k_x=0$. The next TM$_1$ mode is 0.2 meV blueshifted, lying very close to the TE$_0$ mode [intense blue subband in Fig.~\ref{fig:dispersion1}(c)]. The splitting between TM$_0$ and TE$_0$ is 0.36 meV. Higher energy modes, with a weaker degree of polarization, are visible at 1.5407 and 1.5411 eV, marked by orange and black arrows in Figs.~\ref{fig:dispersion1}(b,d) and Figs.~\ref{fig:dispersion1}(c,e), respectively. Analyzing the dispersion relation $k_y$, Figs.~\ref{fig:dispersion1}(d,e), a further, horizontal discretization of the energy levels is clearly observed. The separation in $k_y$ between consecutive antinodes of a single state is $\sim$0.4 $\mu$m$^{-1}$.

\begin{figure}
\setlength{\abovecaptionskip}{-5pt}
\setlength{\belowcaptionskip}{-2pt}
\begin{center}
\includegraphics[width=0.9\linewidth,angle=0]{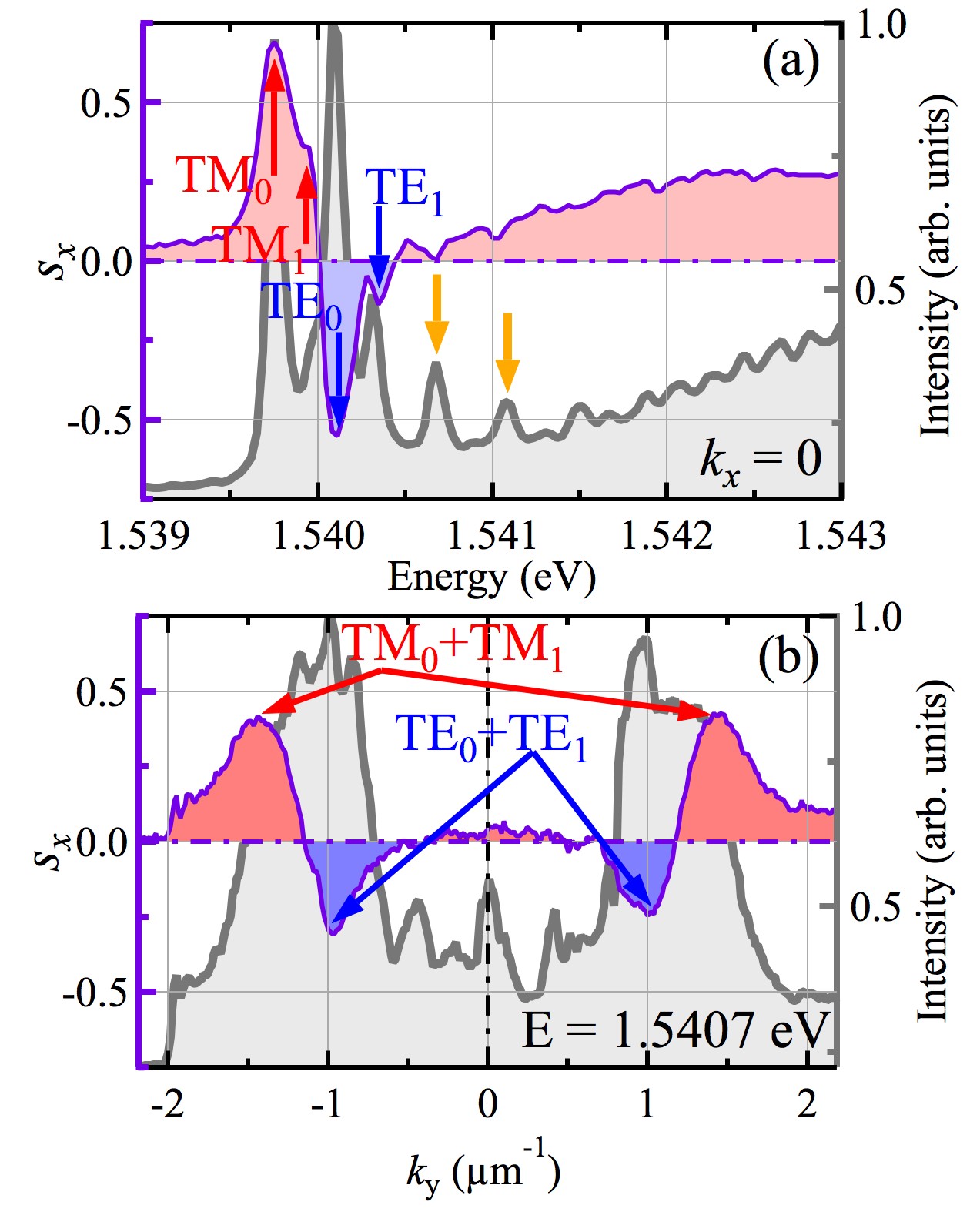}
\end{center}
\caption{(Color online) (a) PL ($s_x$) as a function of energy at $k_x=0$ in a gray, thick (purple, thin) line. (b) PL ($s_x$) as a function of $k_y$ at $E=1.5407$~eV in a gray, thick (purple, thin) line. Red and blue arrows in panels (a,b) mark the TM and TE character of the even subbands in the dispersion relations, respectively. Orange arrows in panel (a) mark the energy positions of weakly-polarized, higher subbands.}
\label{fig:dispersion2}
\end{figure}

Detailed spectra at $k_x=0$, both for intensity (thick, gray line) and $s_x$ (thin, purple line), are given in Fig.~\ref{fig:dispersion2}(a), with different, labelled modes indicated by arrows. The aforementioned higher energy modes are marked by orange arrows. Figure~\ref{fig:dispersion2}(b) details a profile of the PL and $s_x$ versus $k_y$, at $E=1.5407$ eV: the predominant structures at $|0.7|<k_y<|1.8|$ $\mu$m$^{-1}$ in thick gray line are constituted by the modes (TM$_0$+TM$_1$) and (TE$_0$+TE$_1$) modes at high $k_y$ values, while the three central antinodes correspond to other confined modes at lower $k_y$ values. Only when a polarization analysis is performed, TE and TM distributions are resolved (thin purple line), as marked in the figure.

\begin{figure}
\setlength{\abovecaptionskip}{-5pt}
\setlength{\belowcaptionskip}{-2pt}
\begin{center}
\includegraphics[width=0.9\linewidth,angle=0]{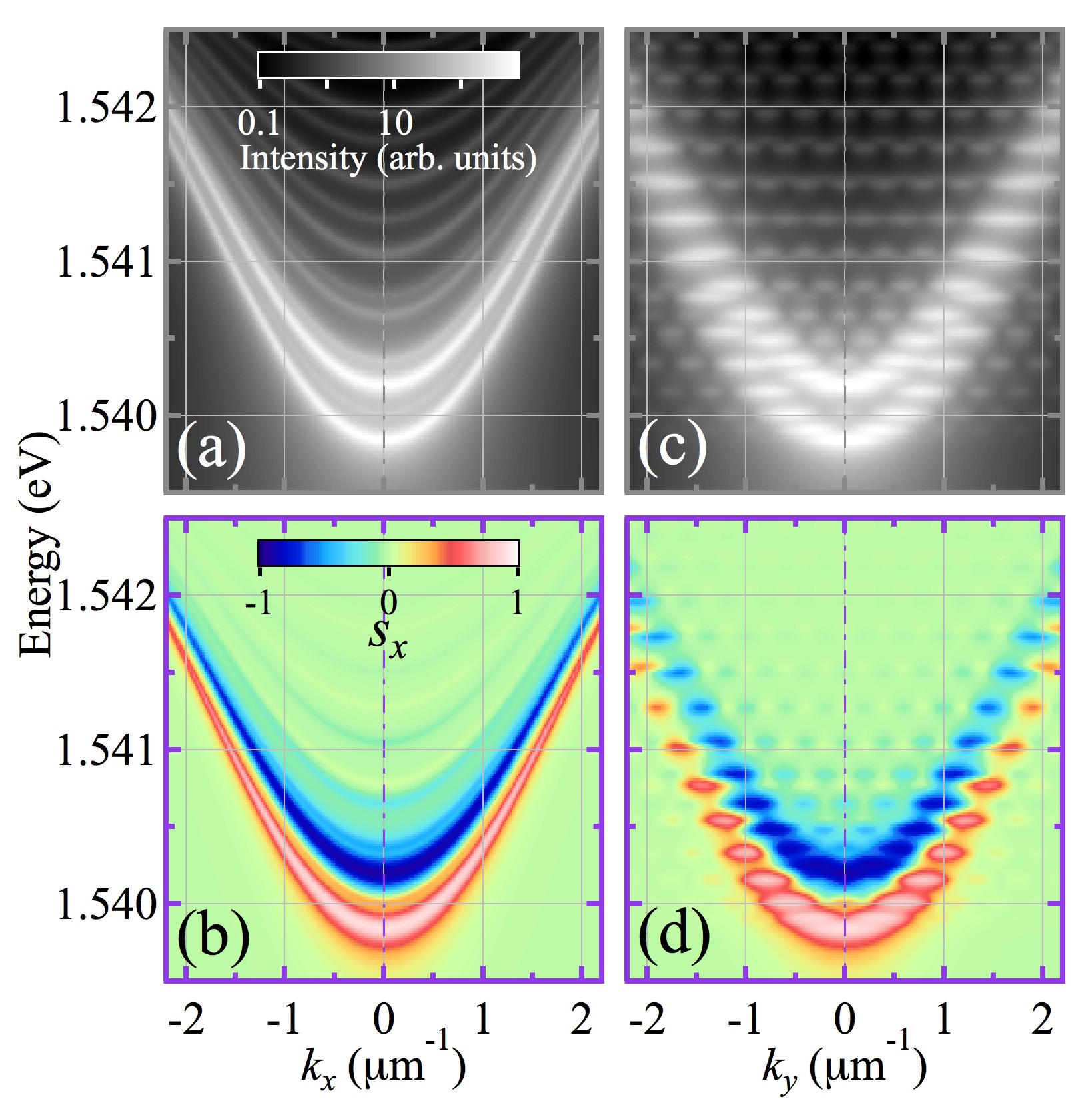}
\end{center}
\caption{(Color online) (a) and (c) Theoretical energy dispersion relation of lower polariton modes along $k_x$ (at $k_y=0$) and $k_y$ (at $k_x=0$), respectively; (b) and (d) corresponding linear degree of polarization ($s_x$). The intensity and $s_x$ are coded in linear, normalized, false color scales.}
\label{fig:teo_dispersion}
\end{figure}

Assuming a square well type potential in the $y$ direction, the energies of the TM- and TE-polarized photonic modes can be approximated by:
\begin{equation}
E_\mathrm{C;TM,TE}(n,k_x)=\frac{(n+1)^2\hbar^2\pi^2}{2m_CL_y^2}+\frac{\hbar^2k_x^2}{2m_C}\pm\Delta_\mathrm{TM,TE}
\end{equation}
where $n=0,1,2...$ is the subband index, $m_C$ is the photon effective mass, $L_y$ is the ridge width and $\Delta_\mathrm{TM,TE}$ characterizes the splitting between the H and V polarizations. The dispersion of the upper ($+$) and lower ($-$) polariton modes is given by the standard two oscillator formula (up to a constant energy shift):
\begin{align}
E_\mathrm{TM,TE}^{\pm}(n,k_x)&=\frac{1}{2}E_\mathrm{C;TM,TE}(n,k_x)\notag\\
&\hspace{10mm}\pm\sqrt{E_\mathrm{C;TM,TE}^2(n,k_x)+4\Omega^2}
\end{align}
where $\Omega=\Omega_R/2$ is the exciton-photon coupling constant. Assuming a Lorentzian lineshape (corresponding to $18$ ps lifetime) and an independent Boltzmann population of the TM and TE polarized energy levels ($T=10$ K), we calculate the dispersions corresponding to lower polariton modes ($E^-$) shown in Fig.~\ref{fig:teo_dispersion}. As in the experiments, we show the $k_y=0$ ($k_x=0$) PL when resolving the dispersion along $k_x$ ($k_y$). The results show that the TM bands are hidden by the stronger populated TE bands at higher energies.
%

\section{Experimental results and discussion}
\label{sec:exp}

Recently, E. Kammann and coworkers reported an analogue of the optical spin Hall effect of an exciton-polariton condensate in a planar MC, under cw, non-resonant, circularly-polarized excitation \cite{Kammann:2012aa}. Circularly polarized condensates propagate over macroscopic distances, while the collective condensate spins coherently precess around an effective magnetic field. \redc{This effective magnetic field can be expressed as $\overrightarrow{H}_{\textrm{eff}}=\frac{\hbar}{\mu_B g} \overrightarrow{\Omega}_k$, where $\mu_B$ is the Bohr magneton, $g$ is the electron $g$-factor, and $\overrightarrow{\Omega}_k$ is the in-plane vector with the following components:
\begin{equation}
	\Omega_x=\frac{\Delta_\mathrm{TM,TE}}{\hbar k^2} (k_x^2-k_y^2), \quad  \Omega_y=\frac{\Delta_{\mathrm{TM, TE}}}{\hbar k^2} k_x k_y,
\label{eq:heff_vector}
\end{equation}
where $\overrightarrow{k}=(k_x,k_y)$ is the in-plane wave vector of the polariton.}

Here we study a similar phenomenon in our quasi-1D structure: we start focusing on the polariton distribution in real space and its degree of circular polarization, under cw, circular excitation, without resolving the PL energy.

\begin{figure}
\setlength{\abovecaptionskip}{-5pt}
\setlength{\belowcaptionskip}{-2pt}
\begin{center}
\includegraphics[width=1\linewidth,angle=0]{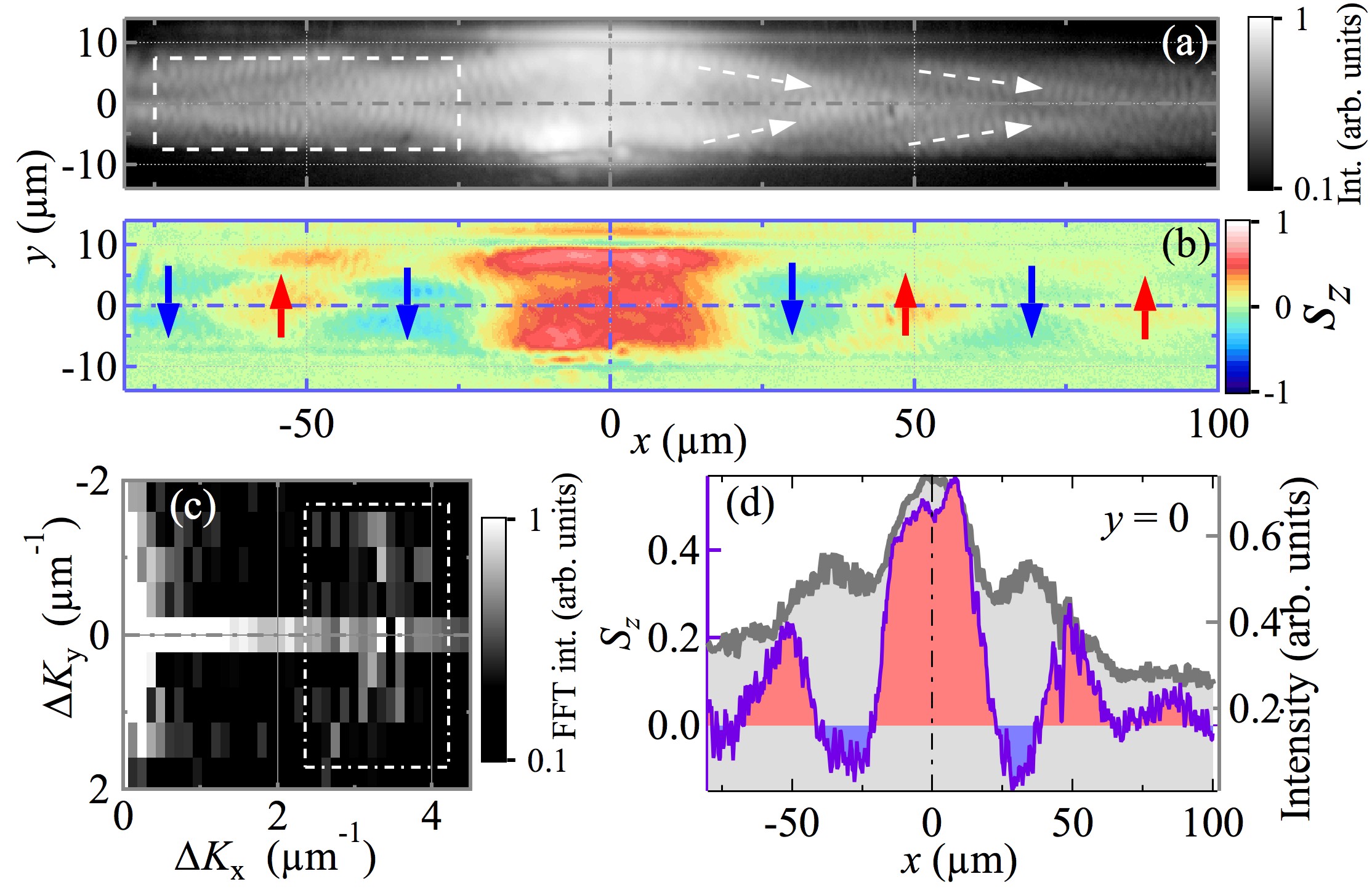}
\end{center}
\caption{(Color online) Collective polariton condensate spin precession in a quasi 1D ridge.  (a) Polariton PL distribution in real space under non-resonant (1.612 eV), circular-polarized ($\sigma^+$) excitation at the center of the ridge. The pump power is $3.75 \times P_{th}$. Dashed white arrows sketch the direction of the polariton flow along its propagation. (b) Corresponding circular degree of polarization distribution ($s_z$). Vertical blue and red arrows highlight the spin precession, oscillating from negative ($\sigma^-$) to positive ($\sigma^+$) values, respectively. (c) FFT intensity of the region enclosed by a dashed, white rectangle in panel (a). The dot-dashed, white rectangle marks the region of relevant frequencies arising from interferences between propagating and backscattered polaritons in real space. \redc{The PL and the FFT map are coded in a logarithmic, false color scale, while a linear one is used for $s_z$.}  (d) PL ($s_z$) versus $x$ in the central region of the ridge $y=0$, plotted with a thick-gray (thin-purple) line.}
\label{fig:int_circ}
\end{figure}

Figure~\ref{fig:int_circ}(a) shows the energy-integrated distribution of the polariton PL in real space, under $\sigma^+$-polarized, non-resonant excitation at $(x,y)=(0,0)$ with a pump power of $3.75 \times P_{th}$. The pump creates outflowing polariton condensates due to the repulsive interactions with the excitonic reservoir.\cite{Wertz:2010ys,Anton:2013aa} The propagation inside the ridge is not purely 1D since slanted traces of the polariton flow are visible (see white dashed arrows as a guide to the eyes), as a result of the reflection of the fluid against the lateral borders at $y=\pm10$~$\mu$m. Interference patterns in the PL, due to polariton-polariton scattering, are also observed (see, for example, the region enclosed by a dashed box). This effect has been also reported in the 2D case.\cite{Christmann:2012hc} A Fourier transform (FFT) of this enclosed region, shown in Fig.~\ref{fig:int_circ}(c), obtains the frequencies corresponding to counter-propagating polariton wave packets, with a difference in momentum propagation of $\Delta K_x\approx 3.4$~$\mu$m$^{-1}$ (see the area enclosed by a dot-dashed box). The corresponding value of $k_x$ matches the typical speed of polariton wave packets in 1D systems ($\sim 1$~$\mu$m/ps).\cite{Wertz:2012ee,Anton:2013aa}

Outside the pump spot, the potential energy is converted into kinetic energy. Polaritons also relax and lose energy through scattering with the excitonic reservoir and through intra-branch scattering;\cite{Wouters:2012aa,Anton:2013aa,Anton:2013ab} the energy of condensed polaritons spans $\sim1.5$ meV across  the subbands (see below). Therefore, the description of the spin distribution in our quasi-1D structure, in the presence of polariton energy relaxation, becomes more complex than in 2D (where the ballistic spin precession occurs in a simpler dispersion relation). However, for the sake of simplicity, we show in Fig.~\ref{fig:int_circ}(b) the energy-integrated distribution of the circularly-polarized component of the PL ($s_z$). The large red area in the central region corresponds to the predominantly spin-up aligned polaritons at the excitation area. The spin of leftwards and rightwards propagating polaritons precesses with a periodicity of $\approx40$ $\mu$m (see up- and down-arrows). The energy integration is responsible for the relatively low values of $s_z$. In Fig.~\ref{fig:int_circ}(d) we quantify both the total PL (thick gray line) and $s_z$ (thin purple line) as function of $x$ at the central cross-section of the ridge ($y=0$). The oscillations in the PL are caused by the fluid reflections against the borders of the ridge, obtaining large intensities when polaritons merge at the center [see arrows in Fig.~\ref{fig:int_circ}(a)]. The spin oscillation and its damping along its propagation are clearly visible (thin purple line). Note that the periodicities of the PL and the spin oscillations do not match since they arise from different phenomena.

\subsection{PL spectroscopy on the spin Hall effect}
\label{subsec:stokes}

\begin{figure}
\setlength{\abovecaptionskip}{-5pt}
\setlength{\belowcaptionskip}{-2pt}
\begin{center}
\includegraphics[width=1\linewidth,angle=0]{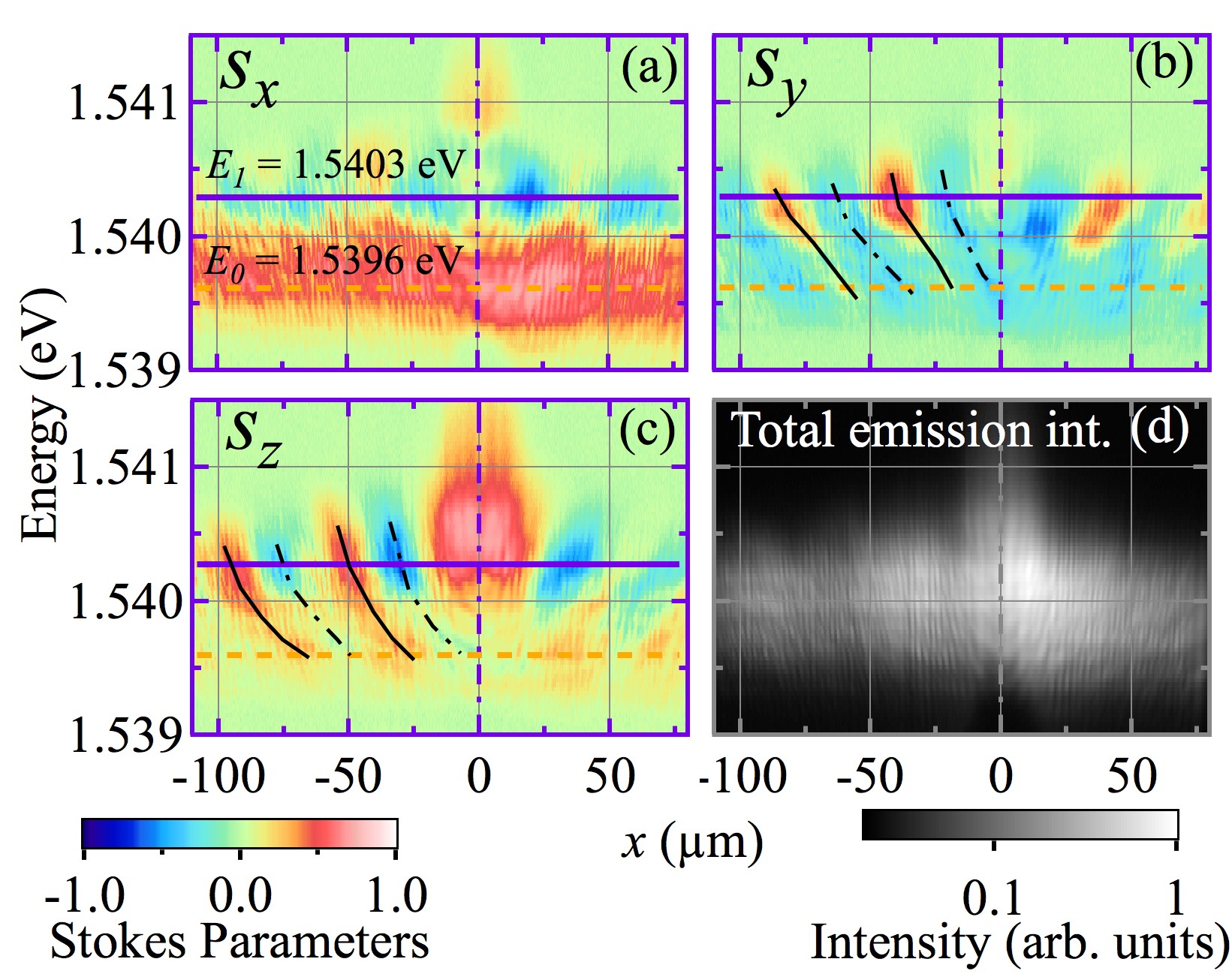}
\end{center}
\caption{(Color online) Stokes parameters of the polariton PL as function of energy and spatial position ($x$): (a) $s_x$, (b) $s_y$, and (c) $s_z$, respectively, under non-resonant (1.612 eV), circular-polarized ($\sigma^+$) excitation. (d) Corresponding PL. The pump power is $3.75\times P_{th}$. Slanted, dot-dashed (solid) lines in panels (b,c) sketch the continuos shift with energy of the minima (maxima) values of $s_y$ and $s_z$. The horizontal purple, solid (orange, dashed) line at $E_1=1.5403$~eV  ($E_0=1.5396$~eV) marks the energy of interest used for the data depicted in Fig.~\ref{fig:prof_stokes}. The PL (degree of polarization) is coded in a logarithmic (linear), false color scale.}
\label{fig:stokes}
\end{figure}

Figure~\ref{fig:stokes} shows the energy- and space-resolved Stokes components of the polarized PL under the same excitation conditions as those described in Fig.~\ref{fig:int_circ}. The polariton condensates span an energy of 1.5~meV around $\sim$1.540~eV. We present here a spatial analysis of $s_x$, $s_y$ and $s_z$ at two different energies $E_0=1.5396$~eV and $E_1=1.5403$~eV, which correspond to those of polaritons condensing into the TM's and the TE's subbands (see in Figs.~\ref{fig:stokes}(a-c) dashed and solid horizontal lines), respectively.
Figure~\ref{fig:stokes}(a) shows a weak spatial oscillation of $s_x$ at $E_1$. Additionally, a small positive $s_x$ from higher-energy excitons (from 1.5405 to 1.5415~eV) is present at $x=0$; this was already present in Fig.~\ref{fig:dispersion1}(c), under below-threshold excitation. At $E_0$, $s_x$ is large and positive, as expected from the TM-character of the lowest polariton subband [see Figs.~\ref{fig:dispersion1}(c) and \ref{fig:dispersion2}(a)]. The diagonal component $s_y$ displays a significant spatial oscillation with a period of $\sim40$~$\mu$m at $E_1$ [see Fig.~\ref{fig:stokes}(b)]. In contrast, $s_y$ barely oscillates around a value of $\sim-0.2$ at $E_0$. Figure~\ref{fig:stokes}(c) shows a highly $\sigma^+$-polarized population at $x=0$ at $E_1$ and above, set by the excitation laser. At $E_1$ the condensed, spreading polaritons exhibit a precessing $s_z$, again with the same period of $\sim40$~$\mu$m. This precession, although weaker, is also seen at $E_0$.

\redc{These oscillations in the Stokes parameters are similar to those previously reported in planar MCs, considering that in our case the propagation takes place along the ridge channel (equivalent to a given radial direction of the 2D rings, see Fig. 3 of Ref. \onlinecite{Kammann:2012aa}). The effective magnetic field, induced by the splitting of the TE-TM modes, is responsible for this precession of the polaritons spin, while they propagate over the ridge, due to the optical spin Hall effect.\cite{PhysRevLett.95.136601} The main difference in our case lies in the energy-dependence of the precession pattern, giving rise to distinct spin textures.} The phase of the spatial $s_z$ oscillations shifts continuously with increasing energies, so that at $E_0$ and $E_1$ they are shifted with respect to each other by a $\pi$ phase approximately. In Figs.~\ref{fig:stokes}(b) and ~\ref{fig:stokes}(c) the slanted, dot-dashed (solid) lines highlight the minimal (maximal) points of the $s_y$ and $s_z$ oscillation across the PL energy, respectively. This phase shift arises from the different propagation speeds of polaritons at different energies: polaritons at higher energies move at higher speeds and therefore travel longer distances for each precession of the spin. Nevertheless, the spin spatial-periodicity does not change significantly with energy. Finally, Fig.~\ref{fig:stokes}(d) displays, for completeness, the PL along the $x$ axis, from 1.5395 to 1.5405~eV.

\begin{figure}
\setlength{\abovecaptionskip}{-5pt}
\setlength{\belowcaptionskip}{-2pt}
\begin{center}
\includegraphics[width=0.85\linewidth,angle=0]{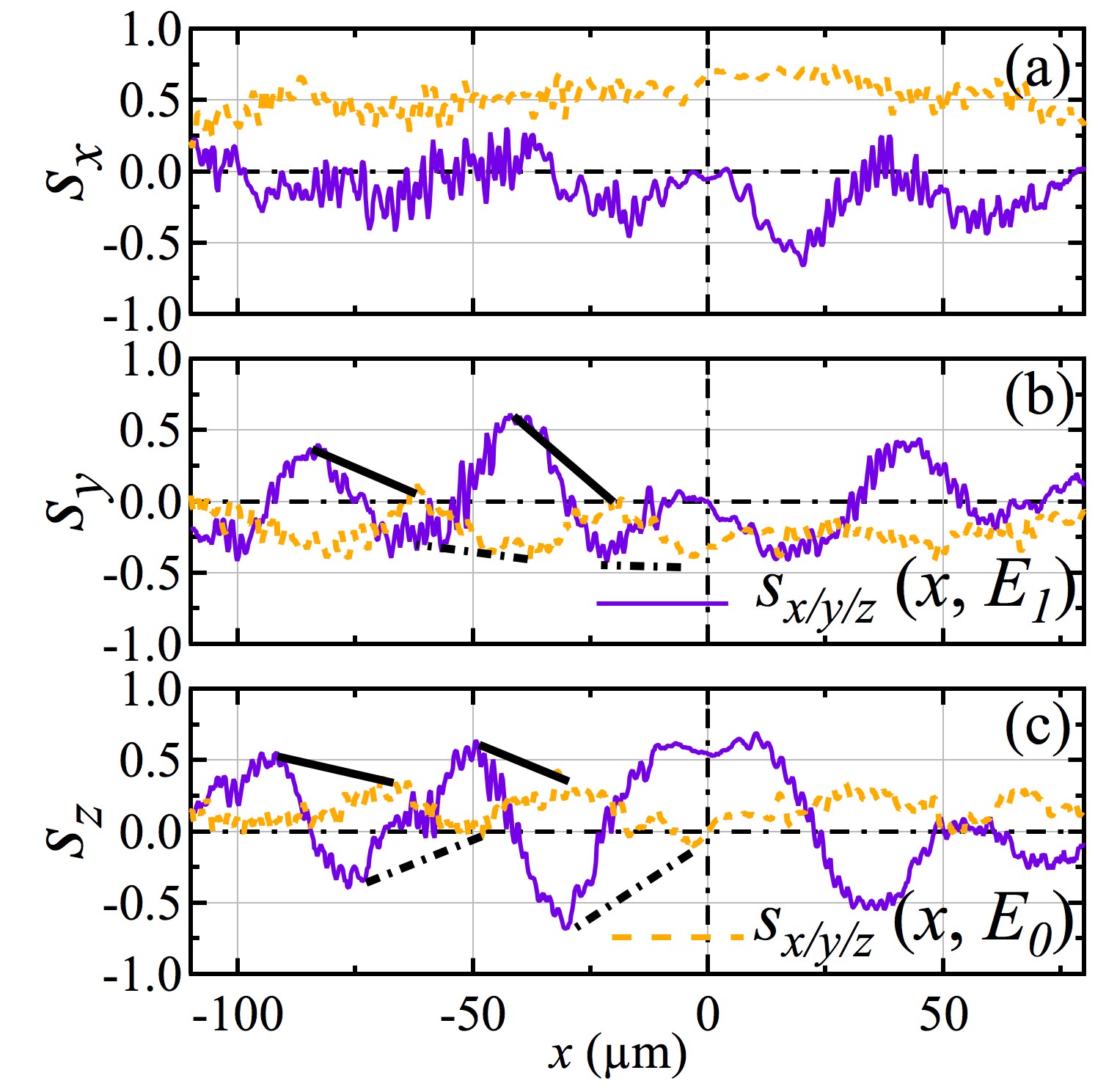}
\end{center}
\caption{(Color online) Stokes parameters of the polariton PL as function of $x$ at energies $E_0=1.5396$~eV (dashed line) and $E_1=1.5403$~eV (solid line), extracted from Fig.~\ref{fig:stokes}. Dot-dashed (solid) lines in panels (b,c) are guides to the eye linking the minimum (maximum) value of $s_y$ and $s_z$ at the two energies, respectively, highlighting the spatial shift with energy of their oscillations.}
\label{fig:prof_stokes}
\end{figure}

In Fig.~\ref{fig:prof_stokes} we detail the different $s_{x,y,z}$ profiles at the two selected energies $E_0$ and $E_1$. The dashed line at $E_0$ shows a constant $s_x\approx0.5$ profile as function of $x$; at $E_1$ (solid line) $s_x$ varies weakly [see Fig.~\ref{fig:prof_stokes}(a)]. Figures~\ref{fig:prof_stokes}(b) and ~\ref{fig:prof_stokes}(c) detail the $s_y$ and $s_z$ oscillations, respectively, at $E_0$ and $E_1$. The displacement in real space of the minimal (maximal) points of the $s_y$ and $s_z$ oscillation from the lower energy $E_0$ to the higher one $E_1$ are evidenced by straight, dot-dashed (solid) lines.

\subsection{Spin precession collapse}
\label{subsec:pump_pow}

A recent study shows that a transfer of the polarization of a non-resonant excitation laser to polariton condensates occurs for excitation powers slightly above the condensation threshold and that the transfer efficiency decays with increasing pump power.\cite{Ohadi:2012aa} We also profit from the former fact to non-resonantly create polariton condensates with a predominant circular polarization. In this section, we investigate not only the latter fact, i. e. the influence of the pump power, but also that of its spin polarization (circular or linear) on the collective polariton spin state and on its propagation.

\begin{figure}
\setlength{\abovecaptionskip}{-5pt}
\setlength{\belowcaptionskip}{-2pt}
\begin{center}
\includegraphics[width=1\linewidth,angle=0]{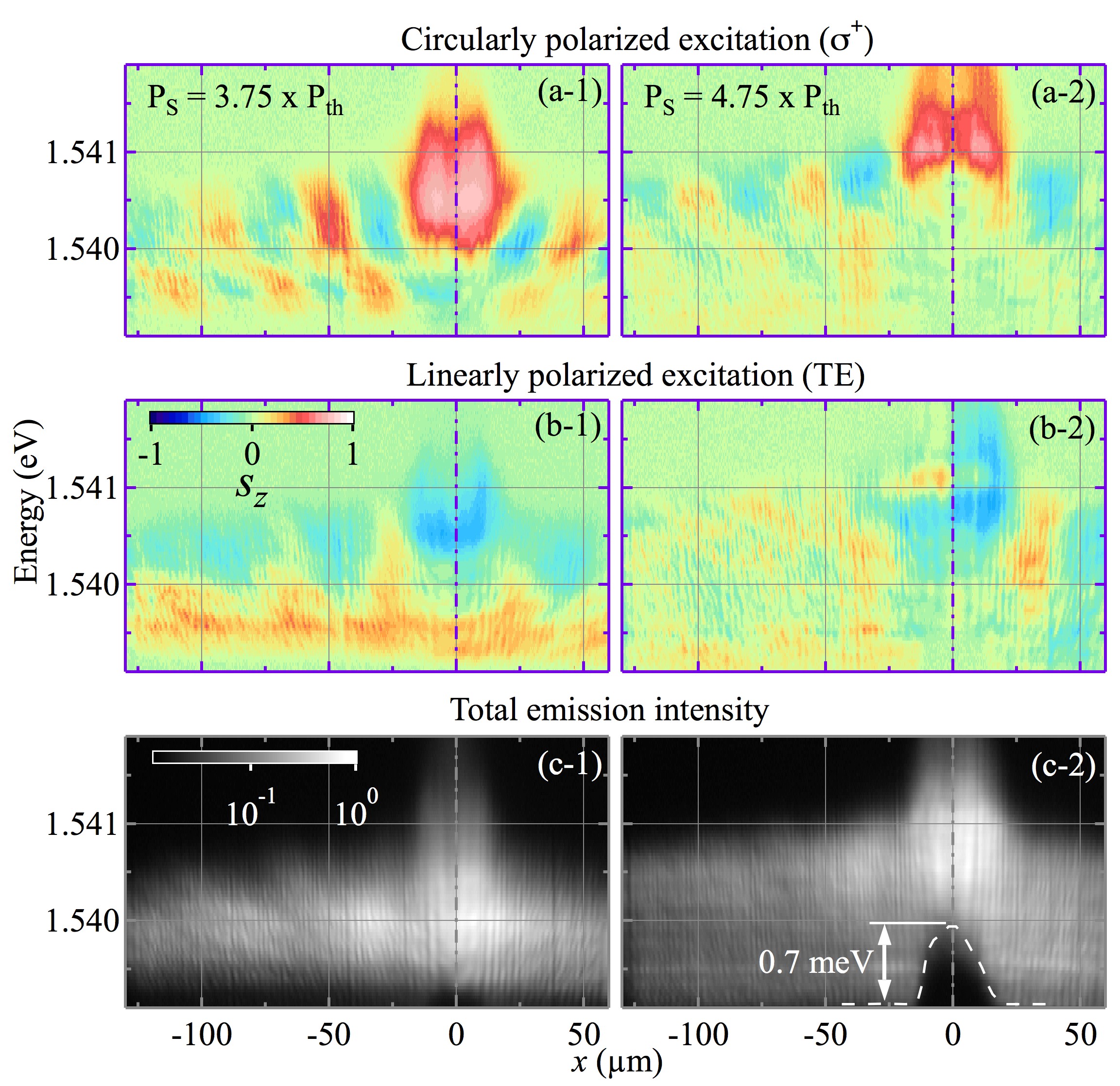}
\end{center}
\caption{(Color online) Polariton spin precession and PL as a function of pump power at $3.75\times P_{th}$ (left column), and $4.75\times P_{th}$ (right column). The non-resonant excitation (1.612 eV) at $x=0$ is circularly- (linearly-) polarized in the first (second) row. Panels (a,b) depict the circular degree of polarization ($s_z$); panels (c) show the polariton PL. In panel (c-2) the local repulsive potential induced by photo-generated excitons at $x=0$ is sketched by a dashed line. The PL ($s_z$) is coded in a logarithmic (linear), false color scale.}
\label{fig:pow_dep}
\end{figure}

Two different pump powers are used for the experiments compiled in Fig.~\ref{fig:pow_dep}: $3.75~(4.75)\times P_{th}$ for the left (right) column. $s_z$ maps as a function of energy and $x$ under $\sigma^+$ (linear excitation) are shown in  Figs.~\ref{fig:pow_dep} (a-1) and ~\ref{fig:pow_dep}(a-2) [Figs.~\ref{fig:pow_dep}(b-1) and~\ref{fig:pow_dep}(b-2)]. Finally, for the sake of completeness, Figs.~\ref{fig:pow_dep}(c-1) and~\ref{fig:pow_dep}(c-2) show the polariton PL.
In Fig.~\ref{fig:pow_dep}(a-1) $s_z$ oscillations are clearly observed from 1.5395 to 1.5408 eV. The $\sigma^+$-polarized, non-resonant excitation induces a highly $\sigma^+$-polarized, blueshifted population at $x=0$, whose PL spans from 1.5400 to 1.5415 eV. A 25$\%$ increase of the pump power strongly reduces the amplitude of the spin precession, which becomes barely visible in Fig.~\ref{fig:pow_dep}(a-2). These oscillations are also suppressed for linear excitation, as shown in Figs.~\ref{fig:pow_dep}(b-1) and~\ref{fig:pow_dep}(b-2). The PL map at high excitation power, 4.75$\times P_{th}$, reveals a non-emitting region around $x=0$ with an energy-width of 0.7 meV and a spatial extent FWHM of $\sim 20$~$\mu$m, highlighted with a dashed line in Fig.~\ref{fig:pow_dep}(c-2). This dark region is caused by the excitonic reservoir, which ejects polariton outwards from $x=0$.

\section{Model}
\label{sec:model}

\begin{figure*}
\setlength{\abovecaptionskip}{-5pt}
\setlength{\belowcaptionskip}{-2pt}
\begin{center}
\includegraphics[width=1\linewidth,angle=0]{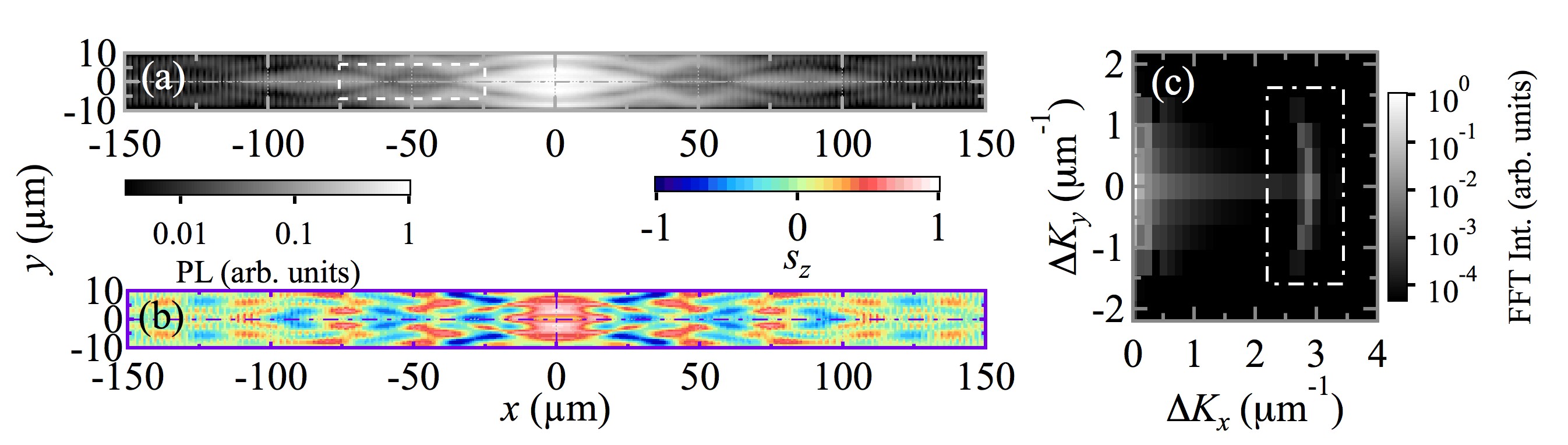}
\end{center}
\caption{(Color online) Simulations of the collective polariton condensate spin precession in a quasi 1D ridge.  (a) PL rendering polariton distribution in real space under non-resonant, circularly polarized excitation at the center of the ridge\redc{, with a pump power $P=3.75\times P_{th}$. The dashed, white box marks the spatial region that is Fourier transformed.} (b) Corresponding simulation on the circular degree of polarization distribution ($s_z$). \redc{(c) FFT intensity of the simulated polariton PL distribution in the framed area in panel (a); remarkable Fourier frequencies arise at $\sim3$ $\mu$m$^{-1}$ from the counter propagating polariton populations, see the delimited region by a dot-dashed, white box. The PL and the FFT map are coded in a logarithmic, false color scale, while a linear one is used for $s_z$.}}
\label{fig:teo_int_circ}
\end{figure*}

To model the spatial structure of polariton condensates we use a mean-field description including incoherent pumping and decay~\cite{Wouters:2007aa} as well as energy relaxation~\cite{Wouters:2012aa}. This model was used previously to describe the dynamics of condensate transistors in microwire ridges\cite{anton:261116,PhysRevB.89.235312}. In the current experiment it is important to use a 2D model that accounts for the subband structure reported in Fig.~\ref{fig:dispersion1} as well as a two-component spinor wavefunction to account for the spin degree of freedom. The spinor polariton wavefunction $\psi_\sigma(\vec r,t)$ obeys the dynamical equation
\begin{align}
i\hbar\frac{d\psi_\sigma (\vec r,t)}{dt}&=\left[\hat{E}_{LP}+\left(\alpha_1 -i\Gamma_{NL}\right)|\psi_\sigma(\vec r,t)|^2\right.\notag\\
&\hspace{5mm}+\left.\alpha_2 |\psi_{-\sigma}(\vec r,t)|^2+V_0(\vec r)+V_{\sigma}(\vec r)\right.\notag\\
&\hspace{5mm}\left.+i\left(W_\sigma(\vec r) - \frac{\Gamma}{2}\right)\right]\psi_\sigma(\vec r,t)\notag\\
&\hspace{5mm}+\Delta_\mathrm{TM,TE}\psi_{-\sigma}+i\hbar\mathfrak{R}\left[\psi(\vec r,t)\right].\label{eq:GP_spin}
\end{align}
where $\sigma=\pm$ denotes the two circular polarizations of polaritons. $\alpha_1$ and $\alpha_2$ represent the strengths of interactions between polaritons with parallel and antiparallel spins, respectively. The operator $ \hat{E}_{LP} = -\frac{\hbar^2\hat{\nabla}^2}{2 m_P}$ represents the parabolic dispersion of the LPB (read LP in Eq. \ref{eq:GP_spin}). Here, $\vec r$ is a two-component vector consisting of the real space coordinates lying on the ridge, the origin of this coordinate system being in the center of the ridge.

Polaritons enter the condensate at a rate determined by $W_\sigma(\vec r)$, which is both polarization and space dependent. While the non-resonant laser used in the experiment is polarized, due to the presence of spin relaxation, one does not expect a full polarization of the photocreated hot excitons. Consequently we expect a partially polarized reservoir of excitons to drive the polariton condensates, eventually yielding both possible circular polarizations. The condensation rate for the $\sigma^+$-polarized polaritons from the excitonic reservoir  is given by:
\begin{equation}
	W_+(\vec r) = W_0 e^{-r^2/L^2} \label{eq: cond_rate}
\end{equation}
where $W_0$ is the peak condensation rate and $L$ is the width. \redc{In principle, the spatial profile of the condensation rate includes the effects of exciton dispersion, diffraction and nonlinear repulsion after excitons are excited by the non-resonant laser pump. In practice, the effective mass of excitons is four orders of magnitude larger than that of polaritons and there is very little spreading of the excitons over length scales relevant for polaritons, such that $L$ can be} taken to be the same as the laser pump-spot diameter. The condensation rate for the $\sigma^-$-polarized polaritons is smaller and given by $W_- = \rho W_+$, where $\rho$ is a parameter that is fitted to the experimental results. In this form, the condensation rate is explicitly spin anisotropic.

The spin dependent effective potential experienced by polaritons can be described by:
\begin{equation}
V_\sigma(\vec r)=G_\sigma W_\sigma(\vec r)
\end{equation}
where $G_\sigma$ is a constant representing the strength of forward scattering processes between excitons in the reservoir and in the condensate.

We also consider a spin-independent component in the effective potential, $V_0$, which is the profile potential of the ridge. We assume it to be that of a 2D infinite square well, where the confinement in the $y$ direction gives rise to the subband structure observed experimentally (Fig.~\ref{fig:dispersion1}) and theoretically (Fig.~\ref{fig:teo_dispersion}).

The polaritons decay with a decay rate $\Gamma$. They also experience a nonlinear loss corresponding to scattering out of the condensate. According to estimates in Ref.~\onlinecite{PhysRevLett.100.250401}, $\Gamma_\mathrm{NL}\approx 0.3 \alpha_1$. Once injected, different circular polarizations are also coupled by the linear polarization splitting $\Delta_\mathrm{XY}$ in the system, which can give rise to oscillations between spin components. While in 2D MC the dominant polarization splitting is wavevector dependent, the dominant splitting in polariton channels is due to strain giving an anisotropic lattice constant.\cite{Dasbach:2005aa} A splitting occurs between polarizations aligned parallel and perpendicular to the channel axis, which remains for zero in-plane wavevector [as can be seen in Figs.\ref{fig:dispersion1}(c,e)], and takes larger values than the TE-TM splitting in 2D MCs.

The final term in Eq.~\ref{eq:GP_spin} accounts for energy relaxation processes of condensed polaritons:
\begin{equation}
\mathfrak{R}[\psi(\vec r,t)]=-\nu\left(\hat{E}_\mathrm{LP}-\mu(\vec r,t)\right)\psi(\vec r,t),\label{eq:relax1}
\end{equation}
where $\nu$ is a phenomenological parameter determining the strength of energy relaxation~\cite{Wouters:2012aa,Wertz:2012ee} and $\mu(\vec r,t)$ is a local effective chemical potential that conserves the polariton population. These terms cause the relaxation of any kinetic energy of polaritons and allow the population of lower-energy states trapped between the pump-induced potentials.

\begin{figure}
\setlength{\abovecaptionskip}{-5pt}
\setlength{\belowcaptionskip}{-2pt}
\begin{center}
\includegraphics[width=1\linewidth,angle=0]{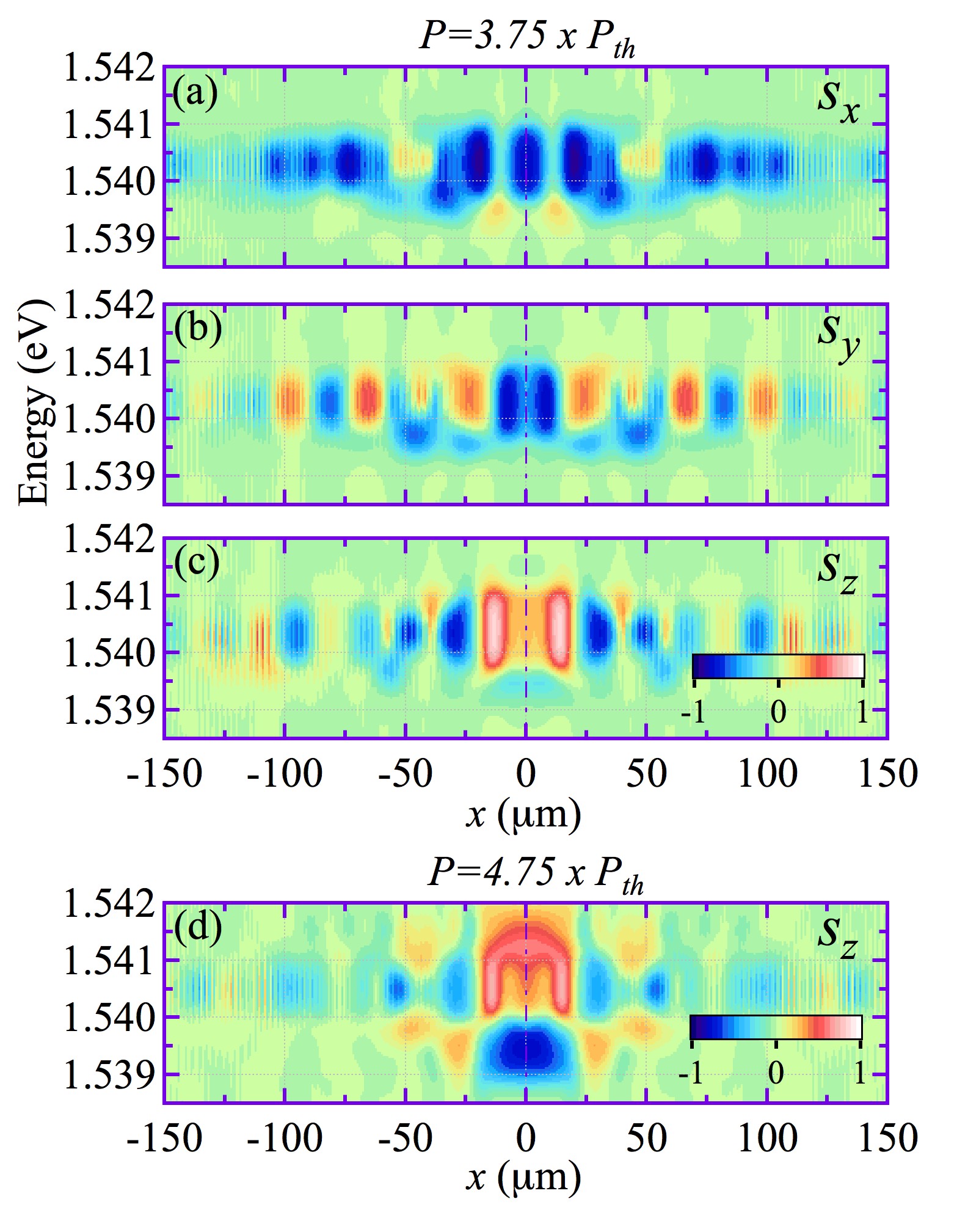}
\end{center}
\caption{(Color online) Simulation on the Stokes parameters of the polariton PL as function of energy and spatial position ($x$): (a) $s_x$, (b) $s_y$, and (c) $s_z$, respectively, under non-resonant, circular-polarized excitation\redc{, with a pump power $P=3.75\times P_{th}$. (d) $s_z$ under a higher pump power excitation $4.75\times P_{th}$.} The degree of polarization is coded in a linear, false color scale.}
\label{fig:teostokes}
\end{figure}

For the simulation that produce the results in Figs.~\ref{fig:teo_int_circ} and \ref{fig:teostokes}, the following parameters are used: $\alpha_1=2.4\times10^{-3}$~meV$\mu$m$^2$, $\hbar\nu=0.14$, $\Gamma= 0.0366$ meV (Ref.~\onlinecite{Anton:2013aa}), $\alpha_2=-0.2\alpha_1$ (Ref.~\onlinecite{Paraiso:2010aa}). The LP dispersion is characterized by an effective mass $m=7.3\times10^{-5}m_e$, fitted to Fig.~\ref{fig:dispersion1}, where $m_e$ is the free electron mass. $G_+=1.0$ and $G_- = 0.7$ are fitted to the measured space- and polarization-resolved energy distributions. $W_0 = 0.185$ meV, $\rho = 0.5$ and $\Delta_\mathrm{TM,TE} = -0.15$ meV. The width of the condensation rate profile is taken to be $L = 10$ $\mu$m. \redc{The calculations were performed using an adaptive step Adams-Bashforth-Moulton error-corrector procedure in a grid with 384x32 points. The polariton wavefunctions were initialized with a weak intensity noise, the distribution of which was found to have no effect on the end result. After a period of initial dynamics, energy distributions are obtained from Fourier transformation over a time window of 250 ps.}

The simulated images shown in Figs.~\ref{fig:teo_int_circ} and \ref{fig:teostokes}(a-c), can be compared to the experimental results in Figs.~\ref{fig:int_circ}(a-c) and~\ref{fig:stokes}(a-c), respectively. \redc{The calculated dependence of $s_z$ on pump power can be observed in Figs. \ref{fig:teostokes}(c,d): a blueshift of the maximum polariton energy as well as a slight reduction of the spin precession with increasing power is obtained, in agreement with the experiments reported in Figs. \ref{fig:pow_dep}(a-1,a-2).} The localized, polarized, incoherent pumping generates two distributions of polaritons separated both in polarization and in energy at $x=0$: while the majority of polaritons are $\sigma^+$-polarized, a significant number of polaritons also condense into a $\sigma^-$-polarized state, which has lower energy due to the spin-dependent blueshifts in the system $V_\sigma(\vec r,t)$. A very good agreement between the experimental and theoretical $(x,y)$ maps is obtained.

The potential $V_\sigma(\vec r)$, which is mostly induced by hot excitons with the same spatial distribution as the pump, represents a strongly repulsive potential in the system that accelerates polaritons outwards. The simulated energy- and space-resolved map of the $s_x$ Stokes parameter, shown in Fig.~\ref{fig:teostokes}(a) is also in reasonable agreement with the experimental results depicted in Fig.~\ref{fig:stokes}(a). As the accelerated polaritons move outward, their spins precess, due to the polarization splitting, giving rise to oscillations in $s_y$ and $s_z$ [see Fig.~\ref{fig:teostokes}(b,c)] as in the experiments [Fig. ~\ref{fig:stokes}(b,c)]. Note that the theoretical model does not reproduce directly oscillations in the spatial distribution of $s_x$. Theoretically, any polarization splitting in the system can always be represented by an effective magnetic field about which the Stokes' vector rotates. It is impossible to find a effective magnetic field that causes oscillations between both negative and positive values in all three components of the Stokes' vector simultaneously (even if multiple forms of splitting are present, the total effective magnetic field cannot make an angle greater than $45^\circ$ with the $s_x$, $s_y$ and $s_z$ axes simultaneously). We thus conclude that the experimentally observed oscillation in $s_x$ is not directly due to spin precession caused by the polarization splitting. Instead, we \redc{speculate} that the oscillations in $s_x$ are linked to the oscillations in the total PL intensity, which competes with a background of incoherent polaritons that are linearly polarized due to the TE-TM splitting. Where the condensate intensity is high, $s_x$ is given by the mean-field theoretical value, while when the condensate intensity is low there may well be incoherent polaritons, not accounted for in the mean-field theory, that give a different polarization. Consequently, oscillations in intensity give the impression of oscillations in the linear polarization degree represented by $s_x$. The oscillations in intensity are due to the 2D nature of the propagation, where both theory and experiment show that polaritons tend to travel at an angle to the $x$ axis, being guided by reflections from the ridge edges. The intensity viewed along the $x$ axis is then greatest when polaritons propagating off-axis cross the $x$ axis.

\section{Conclusions}
\label{sec:conclusions}

In summary, we have studied the optical spin Hall effect in a quasi-1D MC, where the lateral confinement yields a suitable scenario for the intra-branch polariton energy relaxation, enriching the phenomenology of the polariton spin patterns. Thanks to a spectroscopic analysis of the optical spin Hall effect, we have shown that a phase-shift in the oscillations of the $s_y$ and $s_z$ Stokes parameters results from the different speeds of propagation of polaritons. These oscillations collapse either when linear-polarized excitation is used or when the pump power of the circular-polarized excitation exceeds a certain level. Our results are interpreted within the framework of a mean-field model for polariton dynamics, which includes incoherent gain from a polarized exciton reservoir, the energy shift due to the reservoir, TE-TM splitting and energy relaxation. The demonstration of the inversion of the polariton spin as it propagates or relaxes in energy is an important ingredient for realizing polaritonic circuits based on the spin degree of freedom.

\acknowledgments

C.A. acknowledges financial support from the Spanish FPU scholarship. P.S. acknowledges Greek GSRT program ``ARISTEIA" (1978) and EU ERC ``Polaflow'' for financial support. The work was partially supported by the Spanish MEC MAT2011-22997 and EU-FP7 ITN INDEX (289968) projects.


\bibliography{spin_biblio}

\end{document}